\documentclass[fleqn,usenatbib]{mnras}

\usepackage{newtxtext,newtxmath}
\usepackage[T1]{fontenc}
\usepackage{ae,aecompl}

\usepackage{graphicx}	
\usepackage{amsmath}	
\usepackage{amssymb}	
\usepackage{lipsum}

\newcommand{\be}{\begin{equation}}
\newcommand{\ee}{\end{equation}}
\newcommand{\bea}{\begin{eqnarray}}
\newcommand{\eea}{\end{eqnarray}} 
\newcommand{\lcdm}{$\Lambda$CDM}

\title{Debiasing Cosmic Gravitational Wave Sirens} 

\author[R. E. Keeley et al.]{
Ryan E. Keeley,$^{1}$\thanks{E-mail: rkeeley@kasi.re.kr}
Arman Shafieloo$^{1,2}$
Benjamin L'Huillier,$^{1,3}$
Eric V. Linder$^{1,4,5}$
\\
$^{1}$Korea Astronomy and Space Science Institute, Daejeon 34055, Korea\\ 
$^{2}$University of Science and Technology, Daejeon 34113, Korea\\
$^3$ Yonsei University, Seoul 03722 Korea\\
$^4$Berkeley Center for Cosmological Physics \& Berkeley Lab, 
University of California, Berkeley, CA 94720 USA\\ 
$^5$Energetic Cosmos Laboratory, Nazarbayev University, 
Nur-Sultan, Kazakhstan 010000 }

\date{Accepted XXX. Received YYY; in original form ZZZ}

\pubyear{2019}

\begin{document}
\label{firstpage}
\pagerange{\pageref{firstpage}--\pageref{lastpage}}
\maketitle

\begin{abstract}
Accurate estimation of the Hubble constant, and other cosmological parameters, from distances measured by cosmic gravitational wave sirens requires sufficient allowance for the dark energy evolution. We demonstrate how model independent statistical methods, specifically Gaussian process regression, can remove bias in the reconstruction of $H(z)$, and can be combined model independently with supernova distances. This allows stringent tests of both $H_0$ and $\Lambda$CDM, and can detect unrecognized systematics. We also quantify the redshift systematic control necessary for the use of dark sirens, showing that it must approach spectroscopic precision to avoid significant bias. 
\end{abstract}




\section{Introduction}\label{sec:intro} 

Using general relativity (GR) to model the observed waveform of a gravitational wave (GW), the luminosity distance to a GW source can be measured.  This makes GWs from mergers of 
compact objects into standard sirens and offers a potential way to measure the present value of the expansion rate of the Universe, the Hubble constant $H_0$ \citep{Schultz:1986,2005ApJ...629...15H,2006PhRvD..74f3006D}. 

In order to do cosmology with these standard sirens, their redshifts must also be measured. The most straightforward way to obtain the redshift is to use GW systems with electromagnetic (EM) counterpart events (e.g.\ X-ray or optical flashes associated with the merger), where the redshift comes from the EM counterpart.  An alternative to obtain the needed redshift information is to cross-correlate GW events with galaxy redshift surveys, as explored in~\cite{Zhang,2018arXiv180705667F}.  Rather than assigning a redshift and luminosity distance to an individual event, this technique would, in a statistical sense, assign an average luminosity distance to a redshift bin.  These `dark sirens' would allow binary black hole mergers to be used as standard sirens.  Binary black hole mergers are much louder 
(compared to binary neutron star mergers) and so are detectable at much higher redshifts and distances, implying many more of them will be seen. 

The fact that GR is used to calibrate GW standard sirens makes them particularly useful in mapping the cosmic expansion history. The current standard candles used to map the expansion history are Type Ia supernova (SN). On their own, however, SN only measure ratios of distances and so can only constrain the shape of the Hubble distance-redshift relation, not its absolute scale. Thus they require calibration. 

This calibration is currently done with the distance ladder including Cepheid periodic variables and the results of the calibration is summarized as a measurement of $H_0$. 
This Cepheid measurement of $H_0$ has generated significant interest recently since it is currently discrepant with $\Lambda$CDM inferences of Planck measurements of the CMB \citep{2018arXiv180706209P} at the ~4$\sigma$ level \citep{2016ApJ...826...56R,Riess:2019cxk,Joudaki:2017zhq,Keeley19}, potentially pointing to new physics. Since GW do measure an absolute distance scale, they can be used to calibrate the SN distances and estimate $H_0$.  Thus GW standard sirens offer a potentially useful cross check on other methods for determining $H_0$ 
\citep[e.g.][]{2019PhRvL.122f1105F}.

However, as we show in~\cite{Shafieloo:2018qnc}, using overly restrictive model dependent techniques to infer $H_0$ from GW datasets runs the risk of yielding substantially biased results.  This can arise from assuming the acceleration of the Universe is driven by a cosmological constant (the $\Lambda$ in $\Lambda$CDM) rather than being more general, or appropriately agnostic, about the evolution of the dark energy component. For instance if $\Lambda$CDM were assumed, but the dark energy component were truly dynamical (to the extent allowed by current cosmological datasets), then the inferred values of $H_0$ and the matter density $\Omega_{\rm m}$ could be biased at the 3$\sigma$ level.  

At low redshifts, $z\lesssim0.1$, this is less of a problem although a simple linear 
Hubble law is insufficient. Systematics can also arise due to 
peculiar velocities, as well as coherent velocity flows  \citep[e.g.][]{2018arXiv181111723M,2006PhRvD..73j3002C,2006PhRvD..73l3526H}. 
At higher redshift, what we call cosmic standard sirens, such systematics are 
mitigated. Furthermore there is far more volume and a greater number of sources. 
If the redshifts for these sources could be measured (e.g.\ by EM counterparts for binary neutron stars or neutron star black hole mergers, or by cross-correlation in 
the absence of EM counterparts, as with binary black holes \cite{MacLeod2008,Petiteau2008}) and a robust model independent technique shown to 
be effective, then cosmology could be tested to much better accuracy and both 
$H_0$ and the $\Lambda$CDM model could be put to stringent tests. 
The Einstein Telescope \citep{Sathyaprakash10,Zhao11,Taylor12} will certainly be able to detect events out to these high redshifts, though whether enough of these events will have corresponding EM counterparts or are well-localized, remains to be seen.

We quantify the application of model independent statistical techniques to 
accurately and precisely infer $H_0$, and the expansion history of the Universe, 
from mock GW and SN datasets.  In Sec.~\ref{sec:data}, we lay out how we construct 
these mock datasets, aiming for future high precision measurements. We apply 
Gaussian processes in Sec.~\ref{sec:GP} as a model independent method and 
demonstrate its success in reconstructing various expansion histories in an 
unbiased manner. Section~\ref{sec:zerr} addresses the issue of required control of 
redshift estimation systematics, quantifying the effects of both additive and 
multiplicative errors. We conclude in Sec.~\ref{sec:Conclusions}.

\section{Mock Datasets}\label{sec:data}

\begin{figure}
    \centering
    \includegraphics[width=\columnwidth]{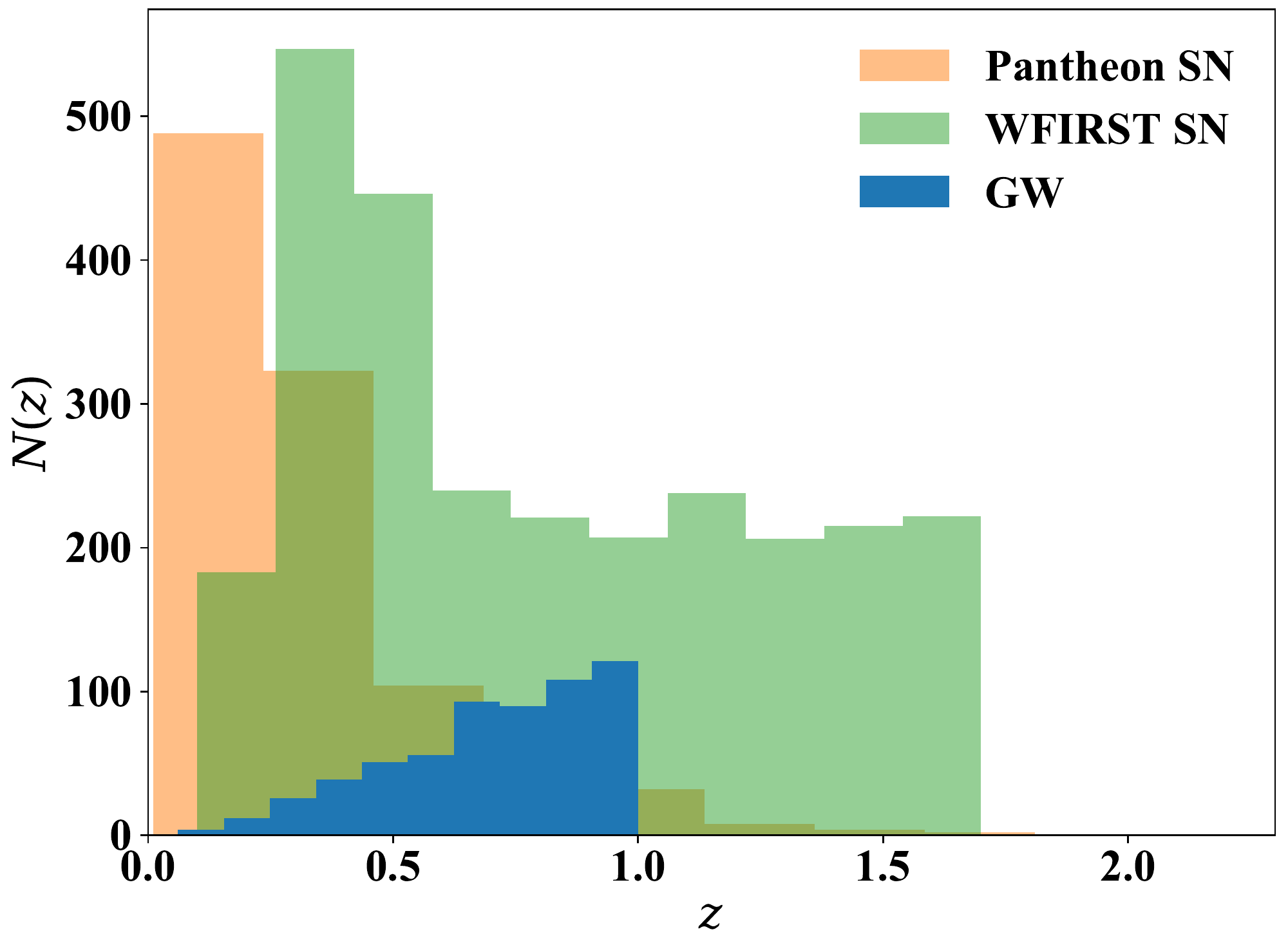}
    \caption{The redshift distribution for GW and SN events in our mock datasets.  The `Pantheon-like' dataset is shown in orange, the forecasted WFIRST dataset is in green, and the next next generation GW dataset is in blue. 
    }
    \label{fig:Nz}
\end{figure}

In order to test how well our model independent methods can recover alternative cosmologies, we generate mock data according to three cosmologies as in \cite{Shafieloo:2018qnc}.  One case is a $\Lambda$CDM cosmology with parameters $H_0 = 69$ km/s/Mpc and $\Omega_{\rm m} = 0.3$.  The other two are dynamical dark 
energy cosmologies with ($w_0, w_a$)=($-0.90,-0.75$) and ($-1.14,0.35$) 
respectively that are consistent with the current joint cosmological probe analysis at the 68\% level in \cite{2018ApJ...859..101S}.  All models have ($H_0$, $\Omega_{\rm m}$) = ($69,0.3$). 

For each cosmology we generate mock GW datasets, `Pantheon-like' SN datasets \citep{2018ApJ...859..101S}, and `WFIRST-like' SN datasets.  For the GW datasets, 
we are interested in how well these model independent methods can do compared to model dependent methods in terms of accuracy.  To this end, we look at the ``Next Next Generation'' case from \cite{Shafieloo:2018qnc}, which has 600 events up to a redshift of $z=1$. This corresponds roughly to a 3rd generation network such as the Einstein Telescope \citep{Sathyaprakash10,Zhao11,Taylor12}. In this optimistic scenario (to test strongly whether bias 
can be overcome out to the maximum redshift), we assume the measured GW source 
redshift distribution follows the cosmic volume element,
\begin{equation}
    \frac{dN}{dz} = \frac{dN}{dV_c} \frac{dV_c}{dz}\ , 
\end{equation}
and we sample from this distribution.  

The redshift distribution for the `Pantheon-like' SN datasets is the same as the actual Pantheon dataset, which includes 1048 SN in the redshift range $0.01<z<2.3$ \citep{2018ApJ...859..101S}.  The redshift distribution for the WFIRST is taken from the WFIRST-AFTA 2015 Report \citep{2015arXiv150303757S}, which forecasts the observation of 2725 SN in the range $0.1<z<1.7$. Each of these redshift distributions is shown in Fig.~\ref{fig:Nz}. 

For the mock GW datasets, the distances are sampled as in \cite{Shafieloo:2018qnc} with a standard deviation of 7\% in distance. This 7\% is chosen to demonstrate a best-case scenario where the precision on $H_0$ is at the level of 1\%.  The distance modulii of the `Pantheon-like' SN dataset are sampled with the covariance matrix of the actual Pantheon dataset. The distance moduli of the `WFIRST-like' SN dataset are sampled with forecasted errors from the WFIRST-AFTA 2015 Report \citep{2015arXiv150303757S}. 

Other studies such as \cite{ZhaoWen18} are less optimistic about the usefulness of future GW surveys.  The authors calculate the average error on the luminosity distance from GW observations as a function of the uncertainty in the position of the event on the sky.  They find that even for the Einstein Telescope the uncertainty on the luminosity distance for an even is around 10\% at $z\sim 0.4$ and the uncertainty increases for larger redshifts.  Networks of Einstein-like Telescopes would yield uncertainties smaller than we use for our mock datasets but would still have a redshift dependence.  This increased uncertainty, coupled with the fact more of the sirens are at larger redshifts for a volume limited sample, reduces the statistical power of the GW dataset.

\section{Gaussian Process}\label{sec:GP}

\begin{figure*}
    \centering
    \includegraphics[width=0.45\textwidth]{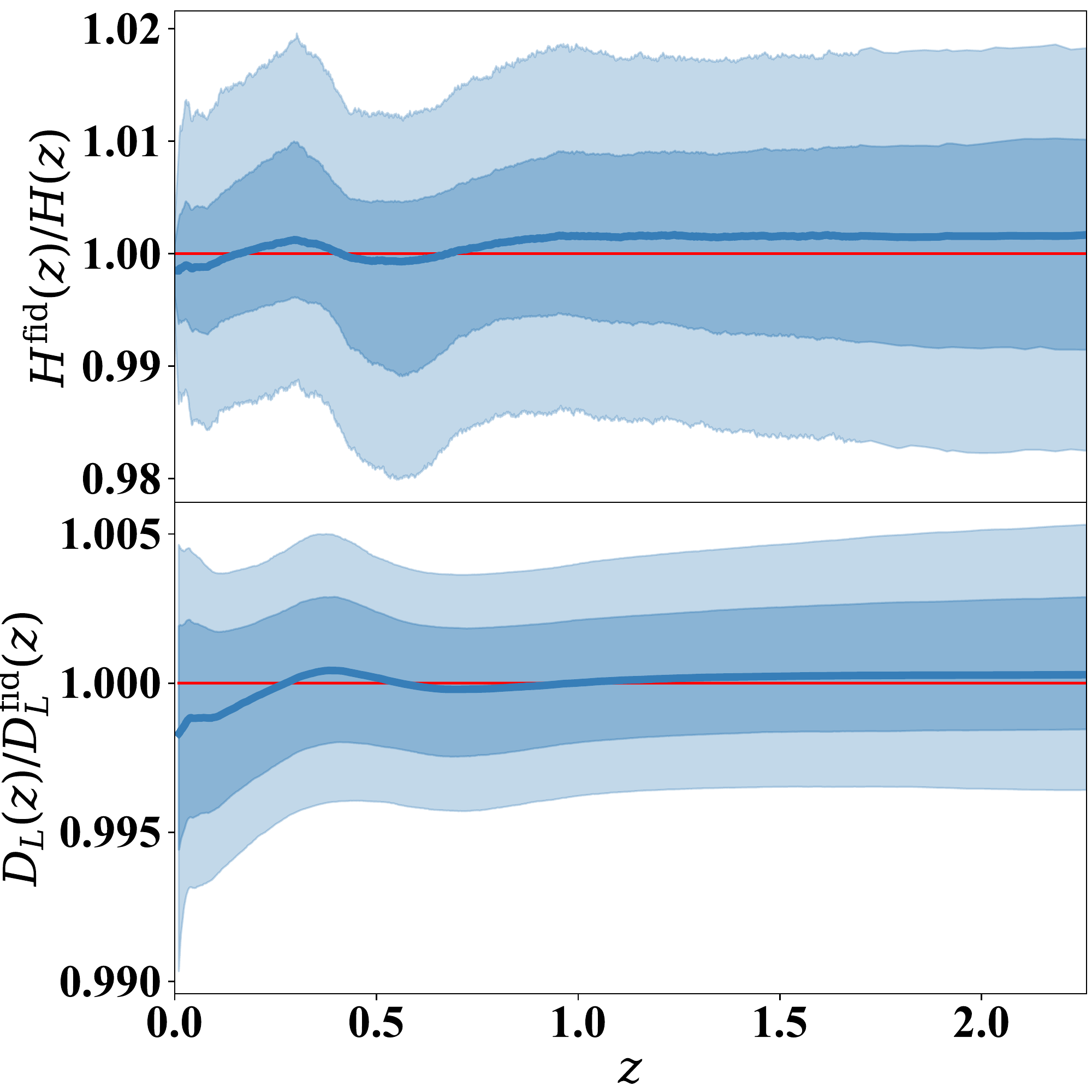}\\ 
        \vspace{0.2cm} 
    \includegraphics[width=0.45\textwidth]{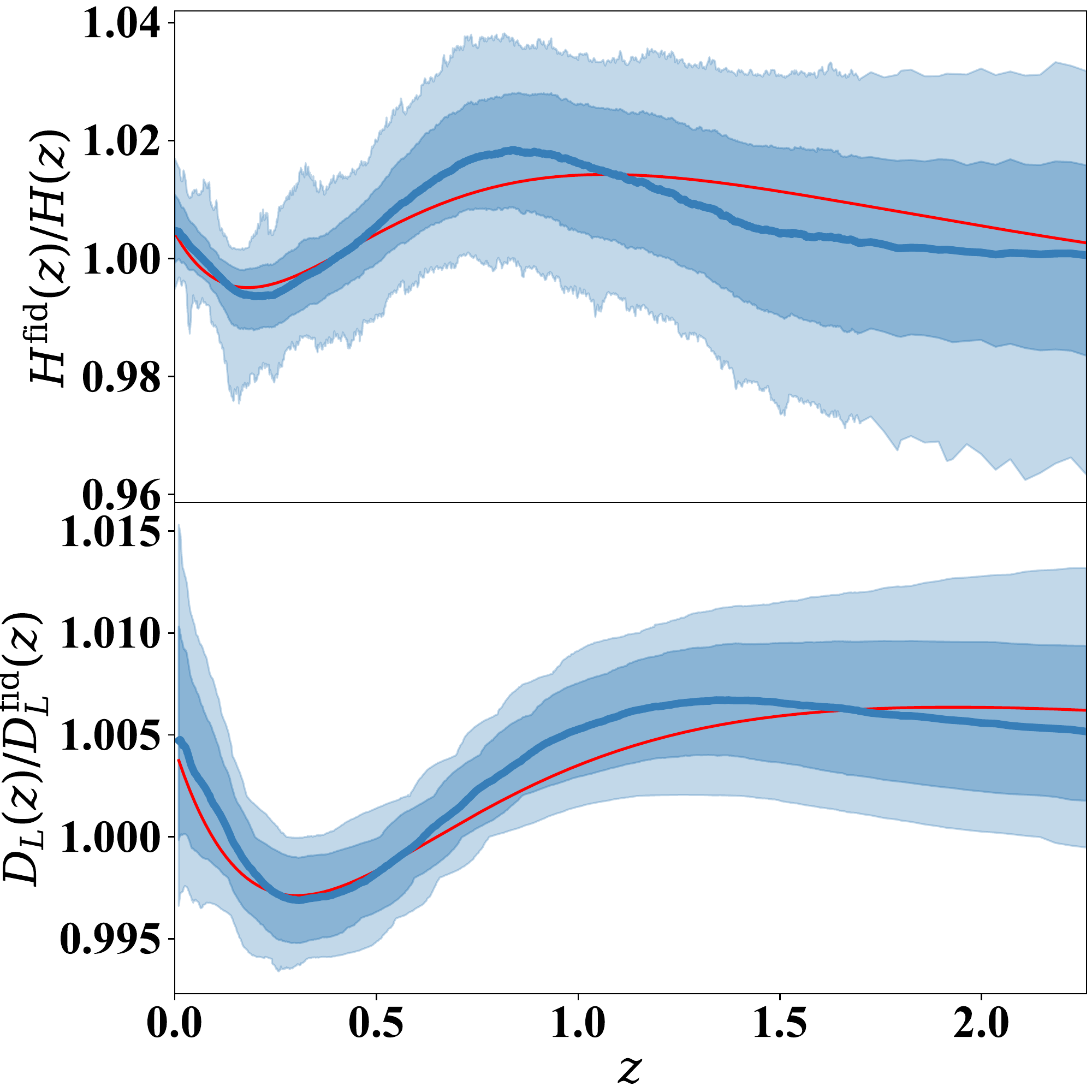}\qquad 
    \includegraphics[width=0.45\textwidth]{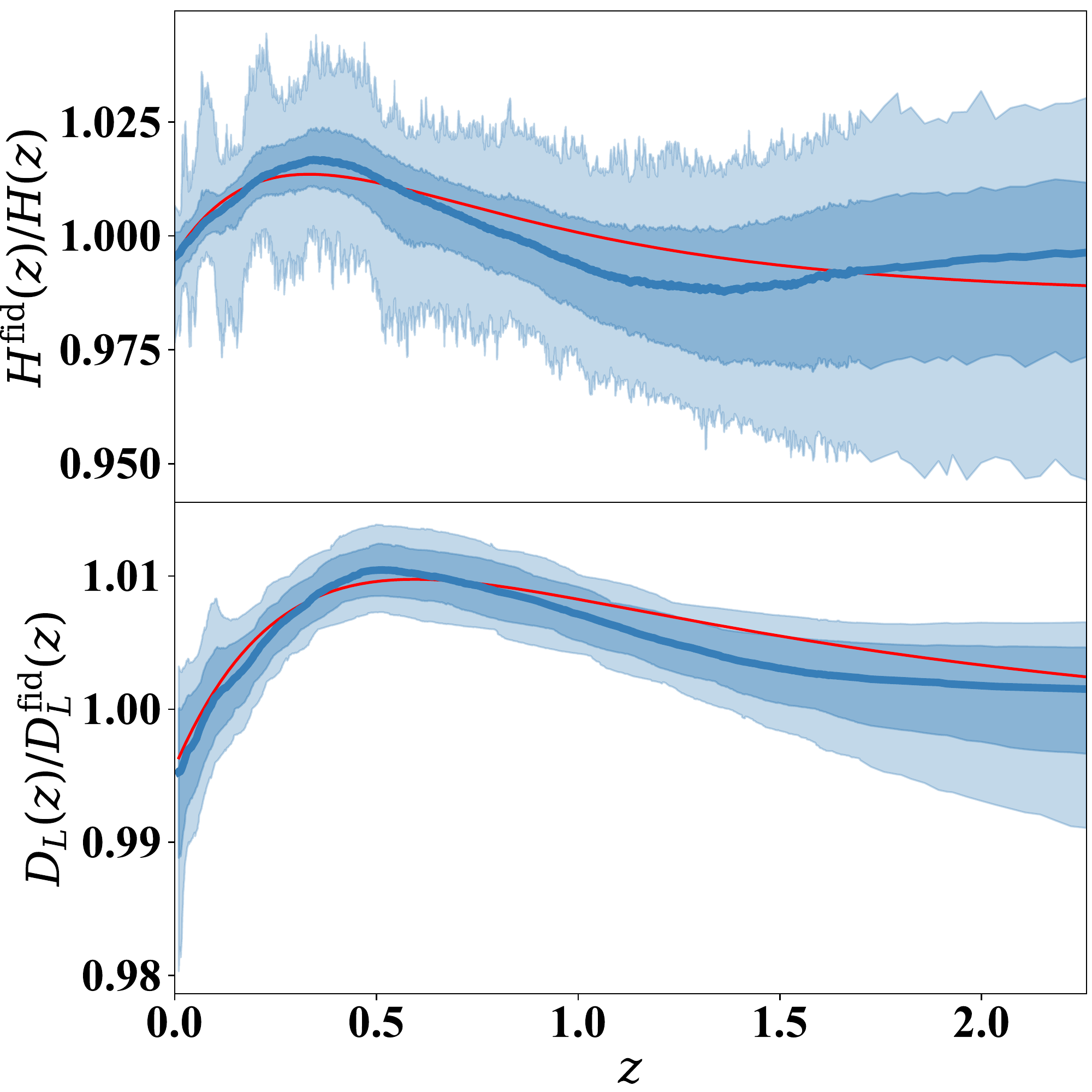}
    \caption{GP reconstructions for the different input cosmologies (top: $\Lambda$CDM, bottom-left: $(w_0,w_a)=(-0.9,-0.75)$, bottom-right: $(w_0,w_a)=(-1.14,0.35)$).  The top panel in each pair is the reconstructed $1/H(z)$ and the lower panel is the reconstructed $D_L(z)$. In both panels the best fit $\Lambda$CDM model is divided out. The thick 
    red curve is the input truth, the thick blue curve is the GP best fit, and the dark and 
    light blue bands are the 68.3\% and 95.4\% confidence intervals. 
    }
    \label{fig:GP_recon}
\end{figure*}

\begin{figure*}
    \centering
    \includegraphics[width=0.45\textwidth]{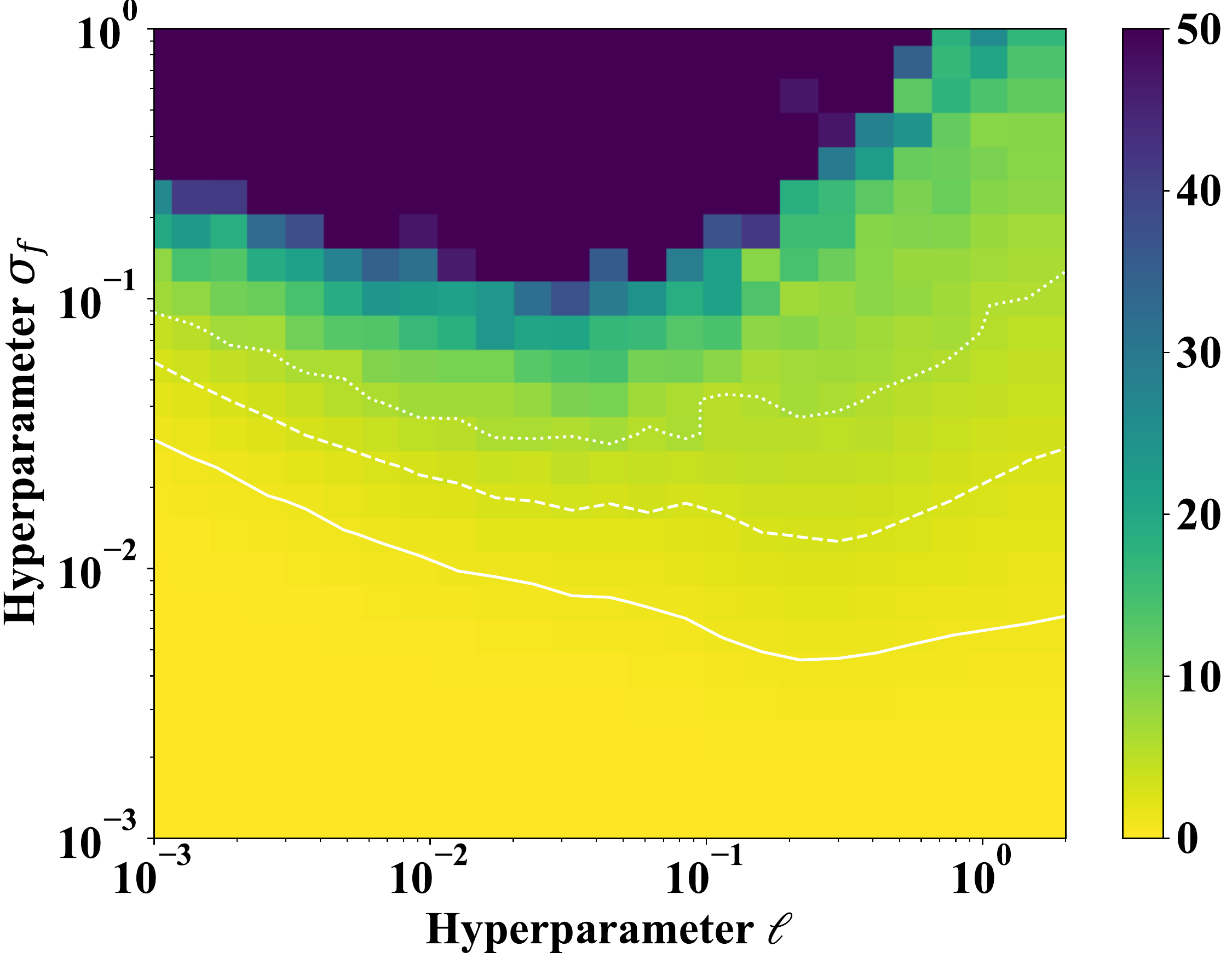}\\ 
        \vspace{0.2cm} 
    \includegraphics[width=0.45\textwidth]{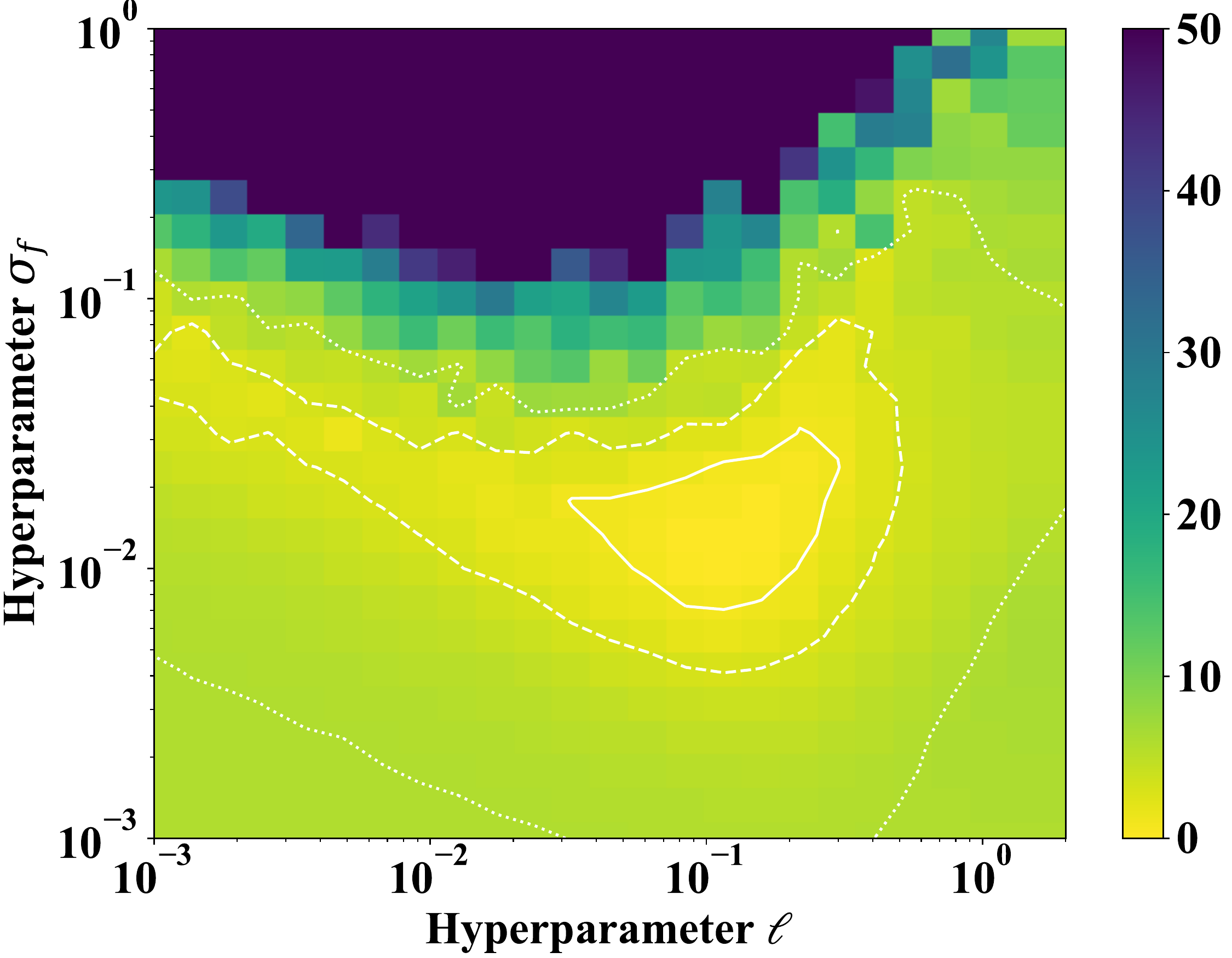}\qquad 
    \includegraphics[width=0.45\textwidth]{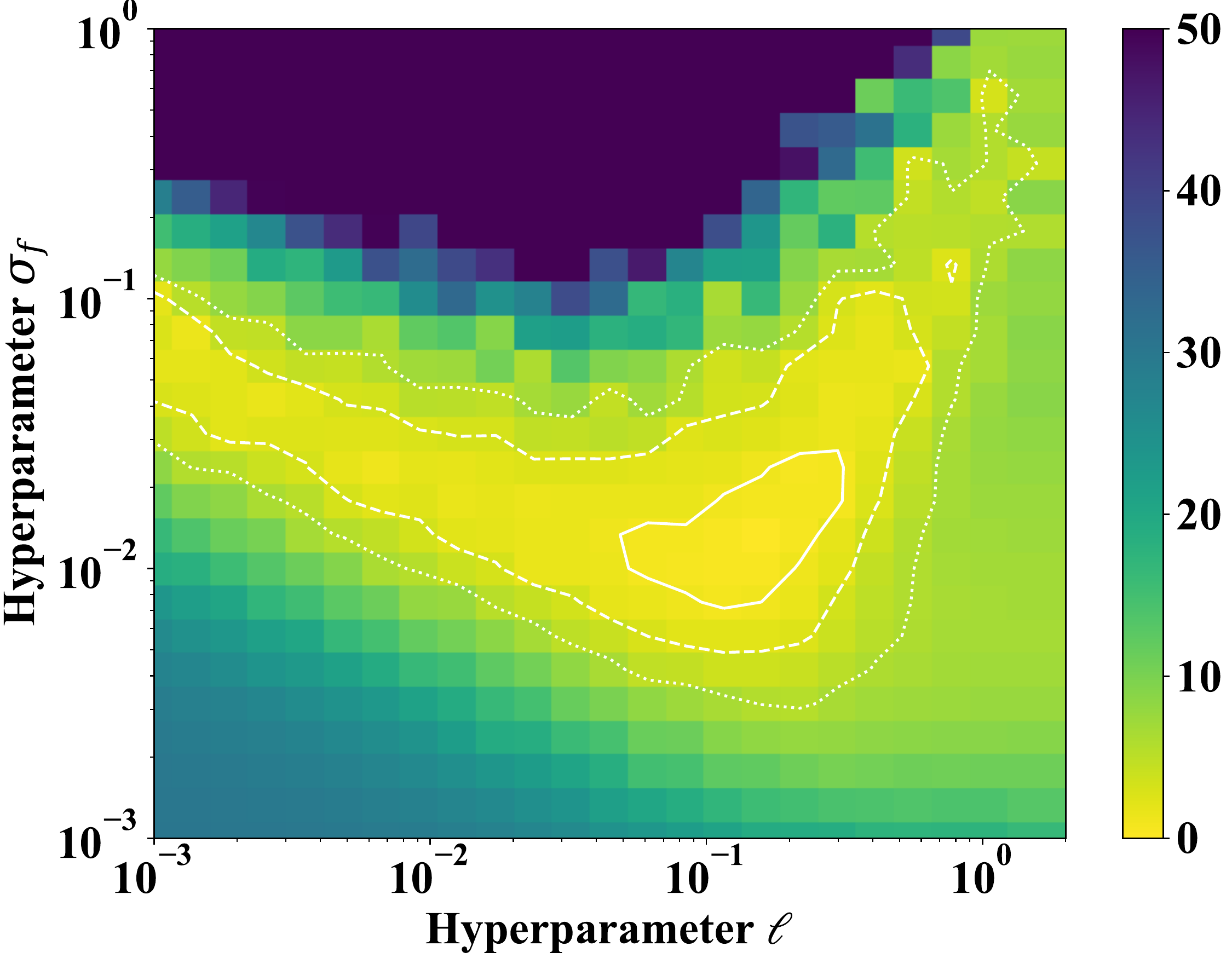}
    \caption{Posteriors of the hyperparameters for the different input cosmologies (top: $\Lambda$CDM, bottom-left: $(w_0,w_a)=(-0.9,-0.75)$, bottom-right: $(w_0,w_a)=(-1.14,0.35)$). The white contours are the 68.3\%, 95.4\%, 99.7\% 
    (1, 2, 3 ``$\sigma$'') confidence levels.  The color corresponds to the $-\Delta$log-likelihood. 
    } 
    \label{fig:hyper_post}
\end{figure*}

\begin{figure*}
    \centering
    \includegraphics[width=0.45\textwidth]{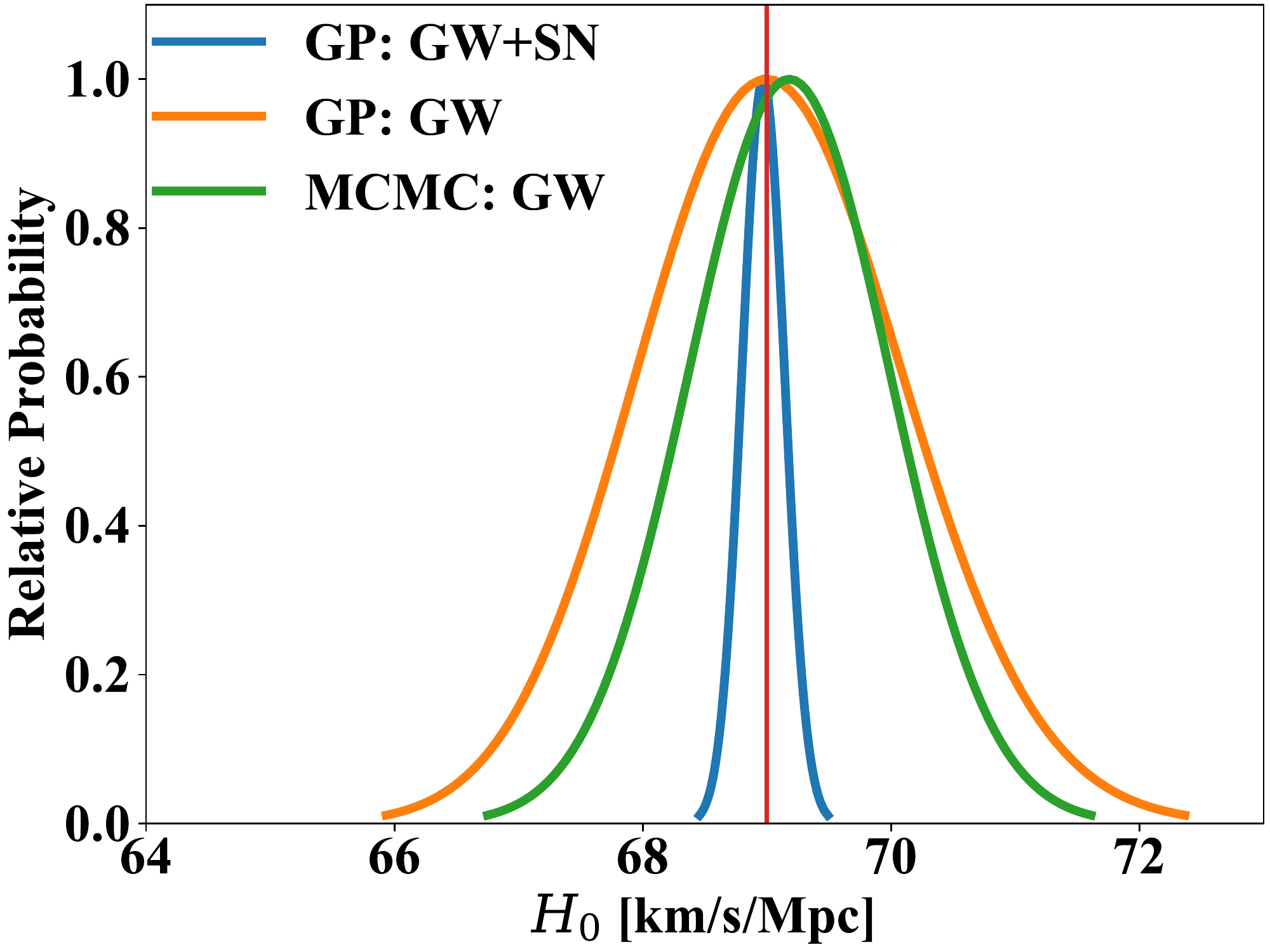}\\ 
    \vspace{0.2cm} 
    \includegraphics[width=0.45\textwidth]{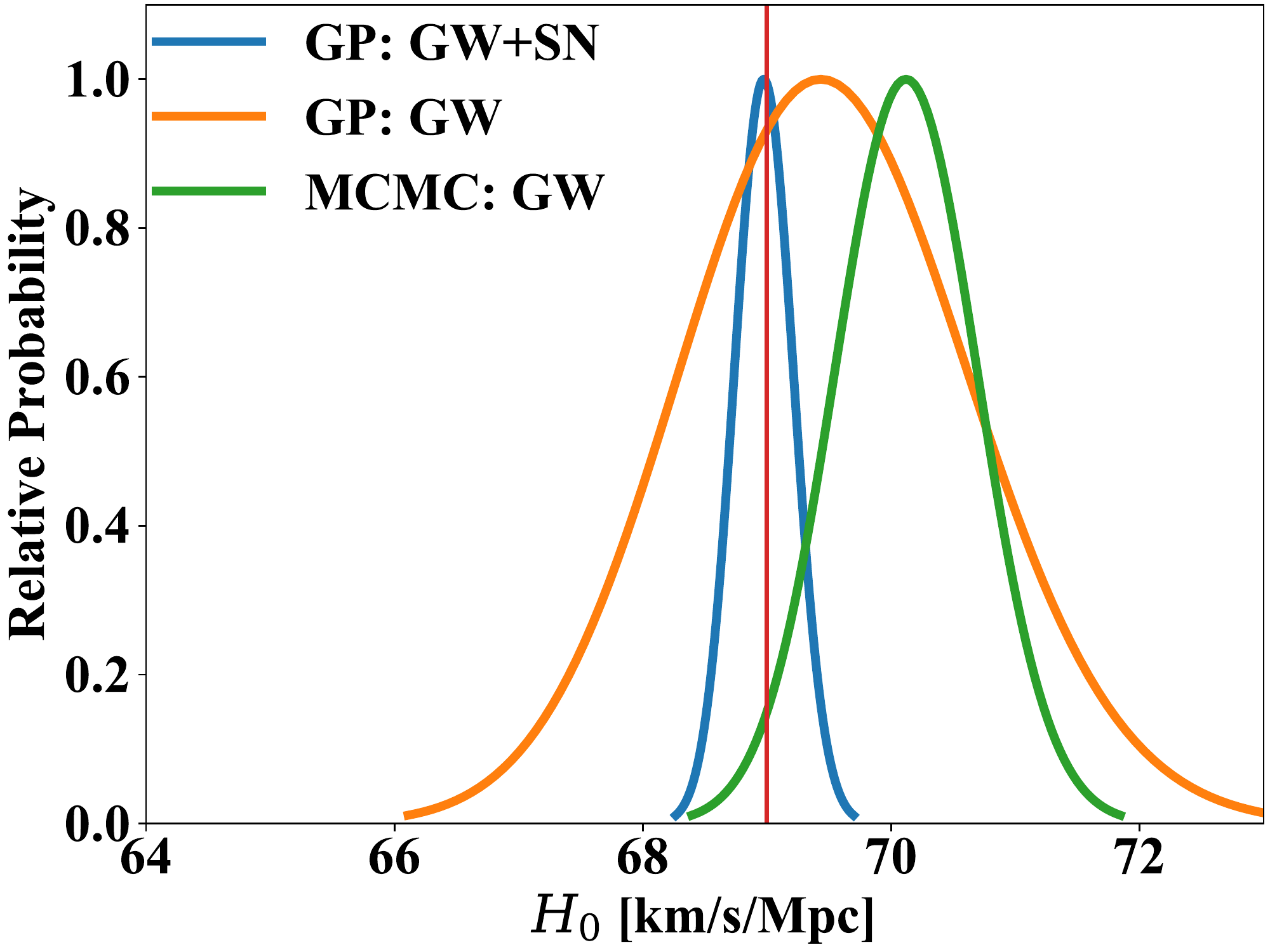}\qquad 
    \includegraphics[width=0.45\textwidth]{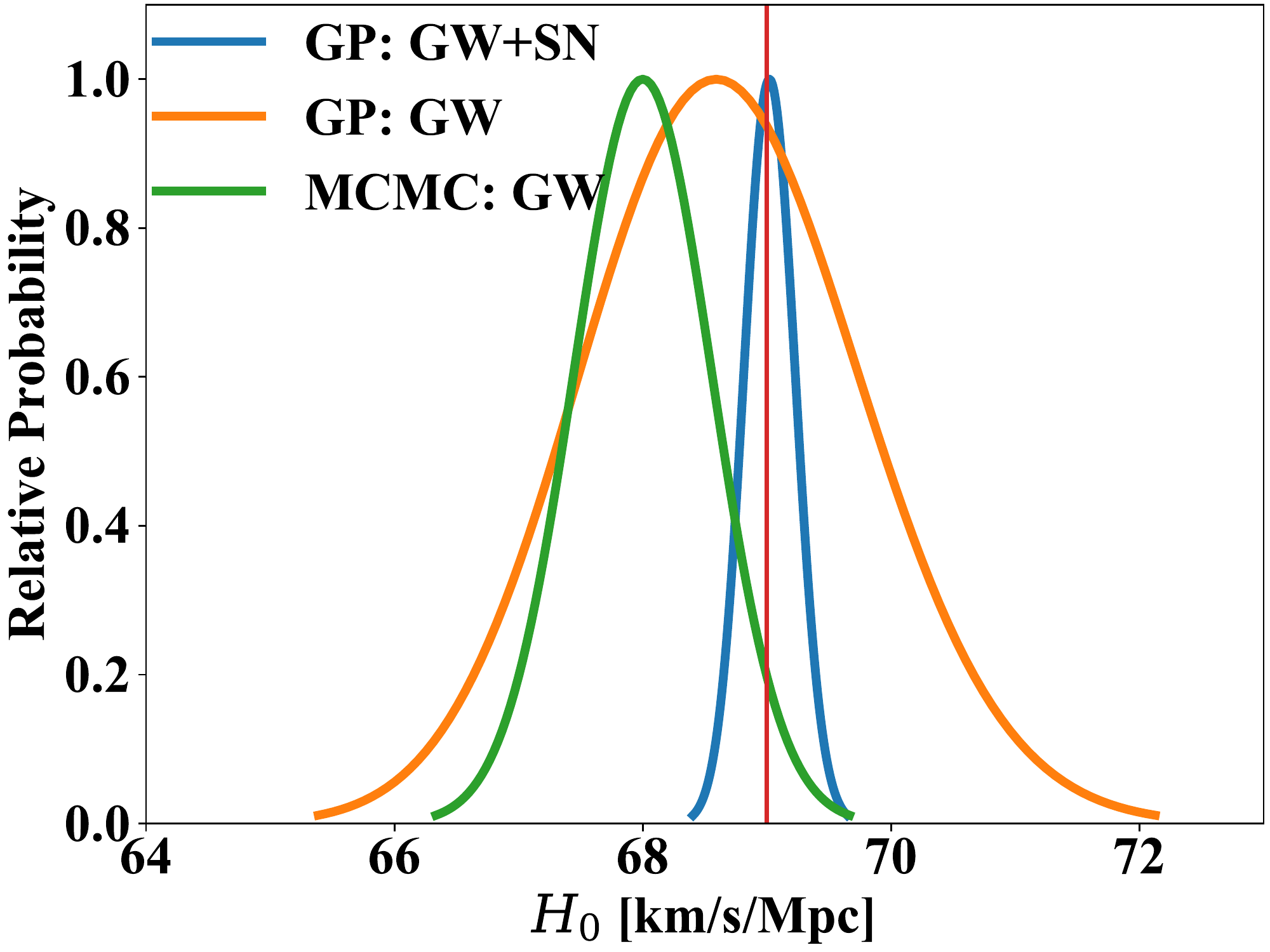}
       \caption{Posteriors of $H_0$ for the different input cosmologies (top: $\Lambda$CDM, bottom-left: $(w_0,w_a)=(-0.9,-0.75)$, bottom-right: $(w_0,w_a)=(-1.14,0.35)$).  The GP posterior for the combined GW and SN datasets is shown in blue and the GP posterior for the GW dataset alone is shown in orange. The MCMC results that assume $\Lambda$CDM (as calculated in \citet{Shafieloo:2018qnc}) are shown in green. The true input value is shown as the vertical red line.}
    \label{fig:H0_post}
\end{figure*}

To infer the expansion history of the Universe in a model independent manner we use Gaussian process (GP) regression \citep{Holsclaw,ShafKimLind2012}. This is a statistical sampling method where instead of sampling a parameter space, the sampling is done over the infinite dimensional space of random realizations of families of curves defined by the GP as informed by the data. In other words, instead of the expansion history being determined by $H_0$, $\Omega_m$, etc.\ and their 
uncertainties, it is determined by a family of model independent curves subject to the GP covariance function between data points. 

For a more accurate reconstruction of $H(z)$ we control the dynamic range by defining 
the variable 
\begin{equation}
    \gamma(z) = \log \left[H_{\rm fid}(z)/H(z) \right]\ . 
\end{equation} 
We use $1/H(z)$ since this is what distances are linearly proportional to. The fiducial 
expansion history $H_{\rm fid}(z)$ is taken to be the best fit $\Lambda$CDM cosmology 
for the given input cosmology (e.g.\ each of our three test models). The log ratio also 
enables a clear test of $\Lambda$CDM -- a deviation from zero points to a deviation from $\Lambda$CDM. 

Since the expansion history is expected to be smoothly varying we use the standard squared exponential covariance function  
\begin{equation}
    \langle \gamma(s_1) \gamma(s_2) \rangle = \sigma_f^2 \, e^{ -(s_1 - s_2)^2/(2\ell^2)}  \ ,
\end{equation} 
where our redshift variable is $s(z) = \log(1+z)/\log(1+z_{\rm max})$, again to control the 
dynamic range. We take $z_{\rm max}=2.3$, the highest redshift of the Pantheon dataset.  
The hyperparameters $\sigma_f$ and $\ell$ play important roles for both error control 
and physical insight, with the first characterizing the amplitude of deviations from the 
fiducial cosmology and the second the correlation scale of fluctuations. Since they have 
physical meaning and impact on the derived cosmology, they cannot be fixed but must be 
fit for. As such, we impose a scale invariant prior on these hyperparameters. Their posterior probability distribution functions carry important information. If $\sigma_f$ 
is consistent with zero this means there is no statistically significant evidence for 
deviations from the fiducial, i.e.\ \lcdm\ cosmology~\citep{ShafKimLind2013,AghHamShaf2017}. If the correlation length $\ell$ is 
very small this may mean one is fitting for noise in the data; if it is very large this 
may mean the data is uninformative about the expansion history.  

The GP regression works by randomly generating a family of functions $\gamma(s)$ described 
by the covariance function, and evaluating the likelihood when comparing to the data. 
As stated, the fiducial cosmologies ($H_{\rm fid}(z)$) are taken to be the best fit $\Lambda$CDM cosmology for each of 
the three cases. 
For the resampled $\Lambda$CDM input cosmology, this is unsurprisingly the input one, namely ($h,\Omega_m) = (0.69,0.3)$.  For the input ($w_0,w_a) = (-0.9,-0.75)$ cosmology, the best fit $\Lambda$CDM cosmology is ($h,\Omega_m) = (0.695,0.286)$, and for the input ($w_0,w_a) = (-1.14,0.35)$ cosmology the best fit  $\Lambda$CDM cosmology is ($h,\Omega_m) = (0.685,0.295)$.

For each of these expansion histories, we then calculate the corresponding luminosity distances, 
\begin{equation}
    D_L(z) = (1+z)\int_0^z dz'/H(z')\ .
\end{equation} 
An advantage of GP is that linear operations (integration or differentiation) on a GP 
are themselves GP. 
The regression is done by weighting these expansion histories by how well their distances fit the data, which is equivalent to calculating the posterior.  The code used for this calculation was adapted from \texttt{GPHist}~\cite{gphistdoi}, which first appeared in \cite{Joudaki:2017zhq}.  This modification is located in an open 
repository\footnote{\url{https://github.com/rekeeley/gphist_GW}}.

The posteriors, i.e.\ the reconstructed expansion histories and distance-redshift 
relations, are shown in Fig.~\ref{fig:GP_recon}.  In both $\Lambda$CDM and $w_0$--$w_a$ example cosmologies, the median of the GP successfully tracks the input values (despite 
using $\Lambda$CDM as the initial, or fiducial, model).  Thus, 
without needing to make any assumptions about the nature of the true expansion history, it can be recovered accurately using GP regression. 

The posteriors of the hyperparameters are shown in Fig.~\ref{fig:hyper_post}.  As discussed previously, these posteriors can be used to determine if the reconstruction is meaningfully different than the mean function.  Specifically, since for each input cosmology we chose the mean function to be the best-fit flat $\Lambda$CDM cosmology for the specific realization of that input cosmology, we can then conclude that if the posterior for the hyperparameters picks out a value for $\sigma_f$ larger than $0$, then these forecasted mock datasets contain information disfavoring flat $\Lambda$CDM. In such a case 
this points to needing some additional physics, e.g.\ dark energy or spatial curvature.  For our input $\Lambda$CDM cosmology, the posterior for $\sigma_f$ is consistent with 0, indicating the data have no preference for anything beyond the best-fit $\Lambda$CDM cosmology.  

However the resampled $w_0$--$w_a$ cosmologies do show the need for flexibility beyond the best-fit $\Lambda$CDM cosmology.  In both cases, the posterior for $\sigma_f$ rules out $\sigma_f=0$ at the $\gtrsim3\sigma$ level.  
In the $(w_0,w_a) = (-0.9,-0.75)$ 
case, $\sigma_f=0$ is ruled out at a moderate significance and in the $(w_0,w_a) = (-1.14,0.35)$, it is ruled out at a more extreme significance.  
Thus, when the true input cosmology includes a dark energy equation of state beyond $w=-1$, this methodology is able to detect the data's preference for information beyond the best-fit $\Lambda$CDM.

Fig.~\ref{fig:H0_post} summarizes the results presented in this paper. They agree with 
the model dependent $w_0$--$w_a$ approach taken in \cite{Shafieloo:2018qnc}.  For each of the different input cosmology cases, the posterior on $H_0$ is shown as 
calculated from an MCMC sampling assuming $\Lambda$CDM (green, as in \cite{Shafieloo:2018qnc}), from a GP regression using only GW data (orange), and from a GP regression using GW and SN data (blue). 
The bias from assuming $\Lambda$CDM is seen most clearly in the $H_0$--$\Omega_{m}$ plane but is seen in just the 1D posterior for $H_0$ as well (green lines).

However, when being more agnostic than assuming $\Lambda$CDM, such as using model 
independent GP regression (orange and blue lines), the bias disappears. We can successfully debias cosmic 
GW sirens, even when they are the sole distance probe. While then 
accurate, even this next next generation dataset will not be more precise than 
$\approx2\%$ on $H_0$ (apart from local sirens). 
If GW are combined with SN datasets -- essentially using GW distances instead of the 
distance ladder to calibrate SN distances  -- then 1\% precision {\it and accuracy\/} 
can be achieved even with an appropriately agnostic model independent method.

\begin{figure}
    \centering
    \includegraphics[width=\columnwidth]{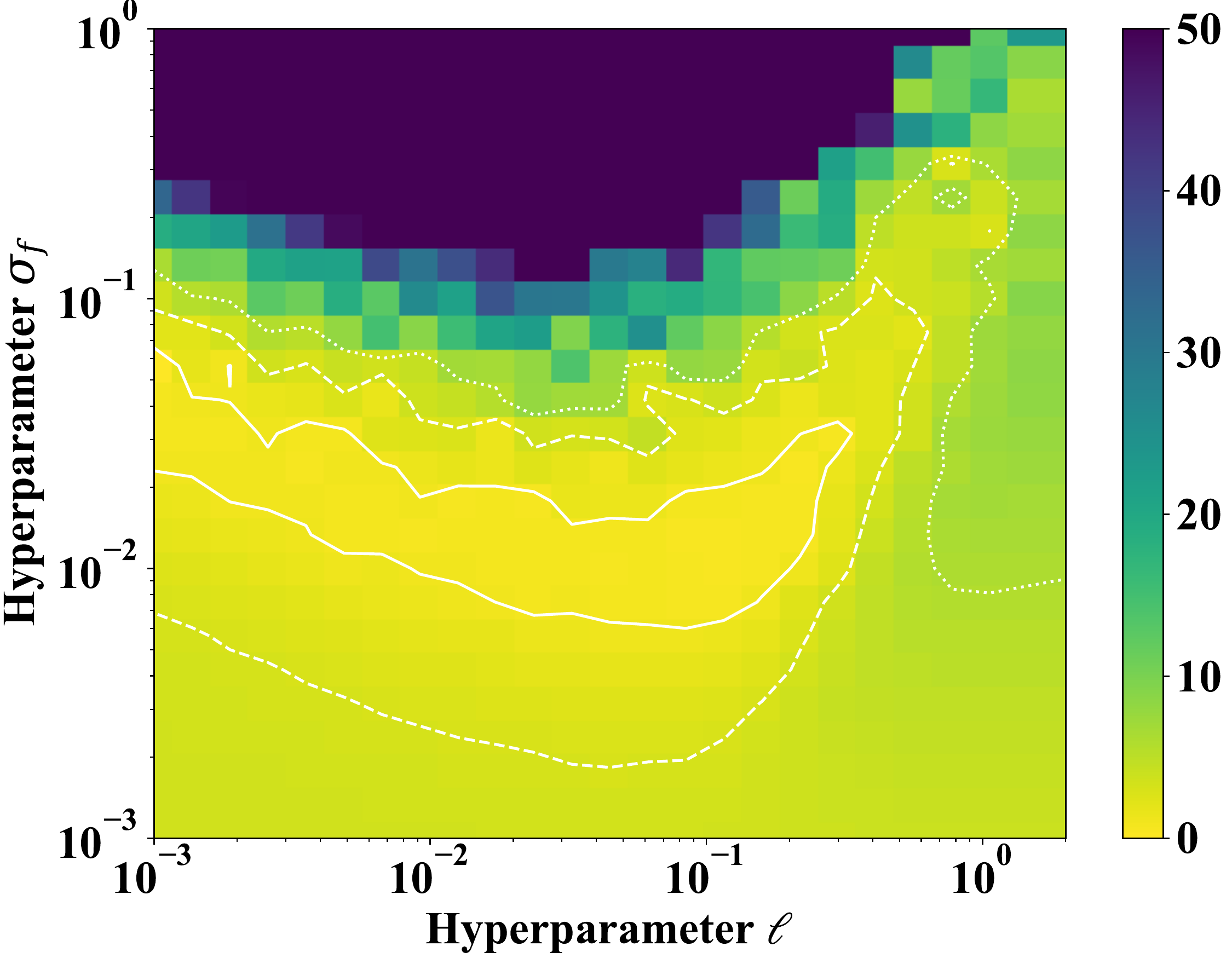}
    \caption{GP hyperparameter posterior for a $(w_0, w_a)= (-1.14,0.35)$ cosmology but now the GW dataset is generated with uncertainties on the distances as in \citet{ZhaoWen18}. The white contours are the 68.3\%, 95.4\%, 99.7\% (1, 2, 3 ``$\sigma$'') confidence levels and the color corresponds to the $-\Delta$log-likelihood.}
    \label{fig:ZWerrors}
\end{figure}

To be concrete about how using less optimistic uncertainties on GW distances such as in \cite{ZhaoWen18}, we repeat our analysis using GW distances as in that paper and see if the GP regression can distinguish the $(w_0, w_a)= (-1.14,0.35)$ cosmology from the $\Lambda$CDM best-fit to the data.  The results are shown in Fig.~\ref{fig:ZWerrors}.  The results are less significant with the less optimistic uncertainties on GW distances.  Whatever mild significance that remains is largely coming from the SN datasets, which are now only loosely anchored by the GW dataset.

\section{Redshift Errors} \label{sec:zerr} 

Dark sirens rely on cross-correlation with large scale structure to estimate the redshift of 
the GW event that should be associated with the measured GW luminosity distance. We now 
examine the accuracy needed for the redshift estimation so as not to bias the cosmological 
parameter determination. 
In particular, a systematic constant offset could look similar to 
a shift in Hubble constant, while redshift dependence might propagate into biases on the 
matter density or dark energy equation of state parameters. 

We begin with a simple redshift residual systematics of the form 
\be 
\delta z=d_0+d_1 z \ ,  
\ee 
i.e.\ an additive and a multiplicative systematic such that 
$z\to(1+d_1)z+d_0$.  
Thus the observable $D_L$ is interpreted as $D_L(z)$ but is really $D_L(z+\delta z)$. We 
can propagate this offset easily into the cosmological parameter estimation through the 
Fisher bias formalism as (see Eq.~3 of \citet{Shafieloo:2018qnc}) 
\be 
\delta p_i=\left(F^{-1}\right)_{ij}\sum_k \frac{\partial{\mathcal O_k}}{\partial p_j}\frac{1}{\sigma_k^2}\Delta{\mathcal O_k}\ , 
\ee 
where the observable ${\mathcal O_k}=D_L(z_k)$, $\sigma_k$ is its 
uncertainty, and $\Delta{\mathcal O_k}$ is the difference between the distance at the assumed redshift and at the true redshift. 

For example, the systematic $(d_0,d_1)=(0.01,0)$ biases the Hubble constant by 
$\delta h=-0.011$ or $1.4\sigma$ and the matter density by $\delta\Omega_m=-0.0195$ or 
$0.8\sigma$ within a $\Lambda$CDM model. While neither of these is too severe, the bias is nearly orthogonal to the degeneracy direction of the 
joint probability contour for $h$--$\Omega_m$, giving a substantial $\Delta\chi^2=75$ 
when fixing to $\Lambda$CDM. Note that this is not purely a shift in $H_0$ because 
$D_L$ is not linearly proportional to redshift for $z\gtrsim0.05$. 

A systematic with some redshift dependence but no low redshift systematic, e.g.\ $(d_0,d_1)=(0,0.01)$, biases $\Omega_m$ 
more substantially, by $2.2\sigma$, and $h$ by $0.6\sigma$, again in a direction such that 
$\Delta\chi^2=44$. Including both systematic contributions, e.g.\ $(d_0,d_1)=(0.01,0.01)$, 
gives a nearly linear additive effect in the parameter biases since they are nearly 
linear proportional to $\delta z$. However, the $\Delta\chi^2$ reacts more extremely since 
it is a product of parameter biases and parameter covariances; for example $(0.01,0.01)$ 
now gives $\Delta\chi^2$=230. 

\begin{figure}
    \centering
    \includegraphics[width=\columnwidth]{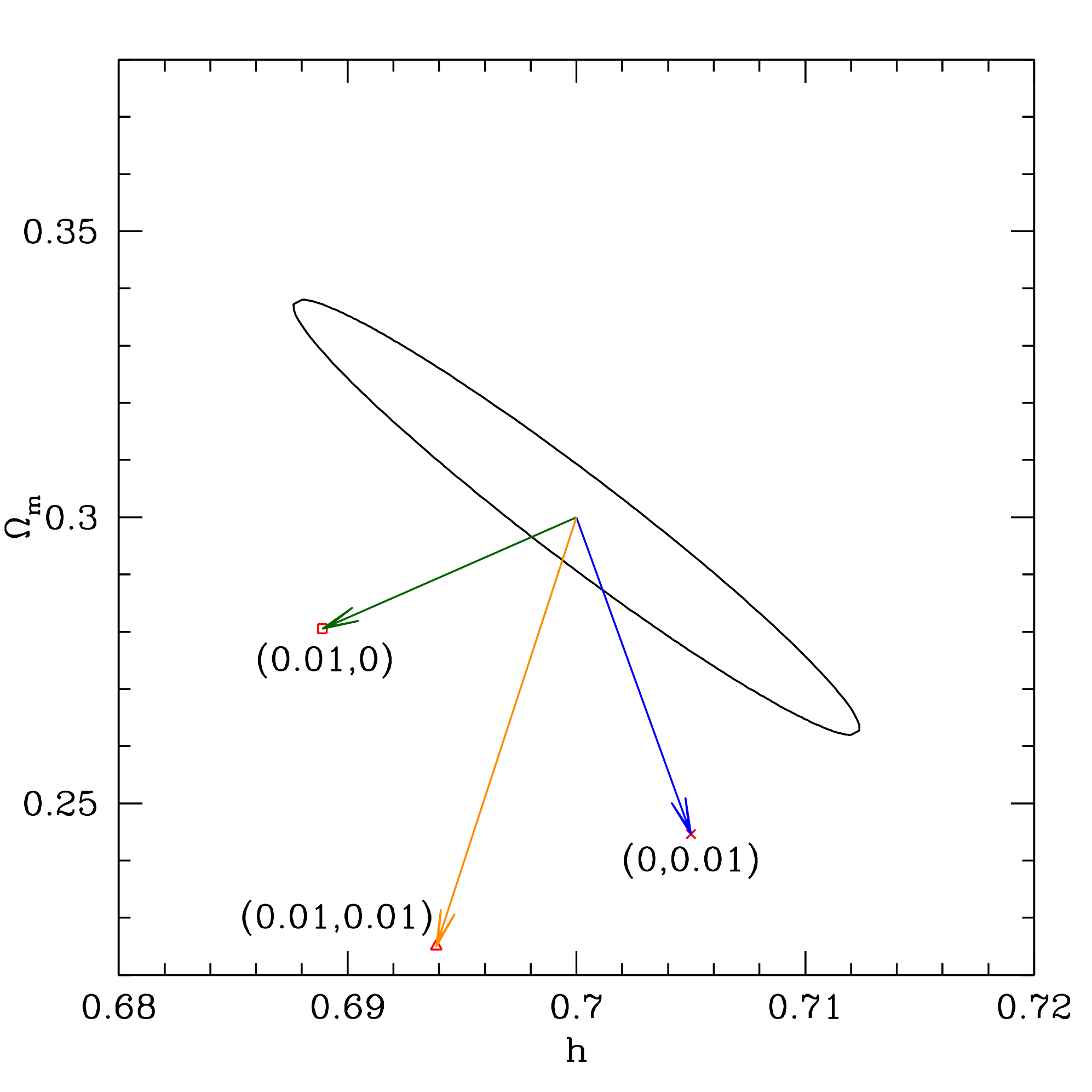}
    \caption{Systematic residuals in dark siren redshift 
    estimation can cause substantial bias in the cosmological 
    parameters. Here we show the statistical 68.3\% CL joint 
    probability contour in the $\Lambda$CDM parameter space, 
    and the bias induced by 
    $\delta z=d_0+d_1 z$, with the square, triangle, square, and x  
    symbols giving the derived cosmological values, labeled by 
    $(d_0,d_1)$.
    }
    \label{fig:homdz}
\end{figure}

Figure~\ref{fig:homdz} shows examples of parameter bias in 
the $\Lambda$CDM model space for various $(d_0,d_1)$. Again note 
that the joint bias in terms of $\Delta\chi^2$ is much larger 
than individual parameter biases, being 44, 75, and 230 for 
the three examples, corresponding to well over $5\sigma$. 

In the $d_0$ only systematic case, we would require the systematics be controlled to 
$|d_0|<0.0018$ to obtain $\Delta\chi^2<2.3$ (i.e.\ $1\sigma$ joint confidence bias). The 
equivalent for $d_1$ only is $|d_1|<0.0023$, and for the more general case of both 
$d_0$ and $d_1$, when they are equal then $|d_i|<0.001$ is needed. This basically 
requires spectroscopic redshift precision for GW sirens to be used as an unbiased 
cosmological probe. Note that the addition of data from other probes, e.g.\ to constrain 
$\Omega_m$ does not help. If we add an external prior of $\sigma(\Omega_m)=0.01$ then 
the statistical errors shrink, and $\Omega_m$ is less biased, but the bias on $h$ can 
actually increase due to covariances. We find the $\Delta\chi^2$ are almost unchanged, 
with the three systematics cases above giving 39, 74, 223 (recall the statistical contour 
shrinks, so even a smaller bias can give a larger $\Delta\chi^2$). 

Returning from Fisher bias to GP regression, we can show how to use GP regression to infer the existence of redshift systematics or other  unidentified systematics (see, for example, 
\citet{BenShafKimLind2019} for the case 
of Malmquist bias or source evolution).  This is done first by calculating the median of the GP inference for one of the datasets (SN for our case), and using this as the mean function in the GP inference for the other dataset.  This allows us to perform the test that if the posterior of the hyperparameters picks out a value for $\sigma_f$ that is significantly above $\sigma_f=0$, then there is some unaccounted for discrepancy between the two datasets.  Since the two datasets are generated from the same Universe, the conclusion would then be that some sort of systematic bias exists in the data.  

To perform an example of this test we use the same mock GW distances and SN distance moduli from a $\Lambda$CDM cosmology, as in the previous section, but the GW redshifts used in the inference are biased by the following equation, $\delta z = \mathcal{N}(0.01,0.01)\,z$, where $\mathcal{N}(0.01,0.01)$ is a normal distribution with mean $0.01$ and standard deviation $0.01$. 
(This is a Monte Carlo version of the $d_1$ case above.) 

\begin{figure}
    \centering
    \includegraphics[width=\columnwidth]{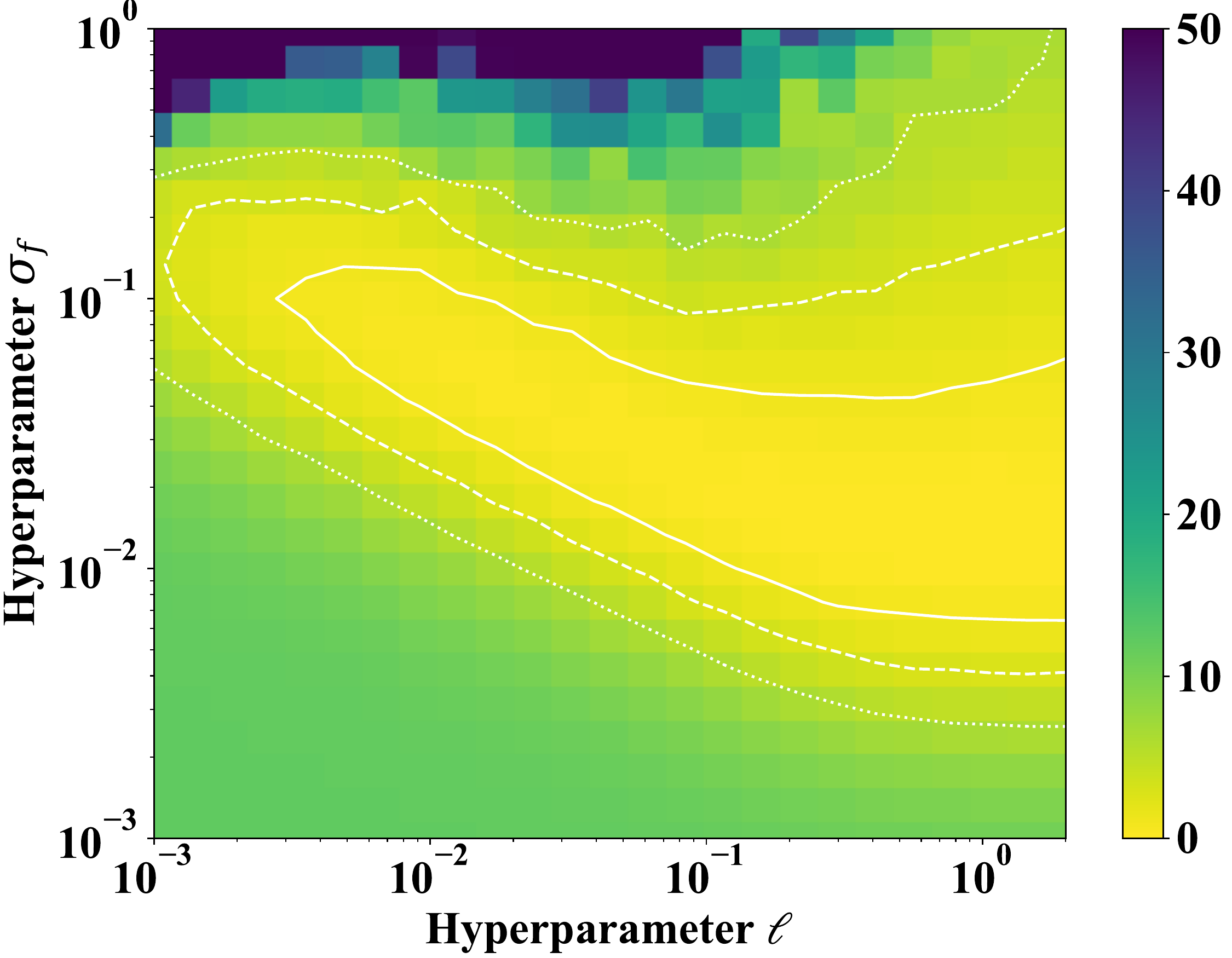}
    \caption{GP hyperparameter posterior for GW data where the mean function of the GP is taken from the median GP inference of the SN data but a GW data redshift systematic exists. The offset from $\sigma_f=0$ indicates discrepancy between the data sets. 
    }
    \label{fig:GPdataxdata}
\end{figure}

The result for this systematics test from biased redshifts in the GW dataset is shown in Figure~\ref{fig:GPdataxdata}.  The posterior of the GP hyperparameters picks out a value for $\sigma_f$ that is significantly above $\sigma_f=0$ (at more than 99.9\% level).  This indicates that the GP regression is able to identify a systematic discrepancy between the GW and SN datasets (it even 
correctly identifies the order of magnitude of the effect).  This test can only identify that some sort of systematic bias exists in the data, not that such offset comes specifically from biased redshift measurements.

\section{Conclusions}\label{sec:Conclusions}

GW sirens are a new distance measure with some 
unique characteristics. They have the potential to contribute 
to mapping the expansion history of the universe, including 
determining the Hubble constant, if appropriately treated within 
the cosmological context. In particular, assumptions about the 
background cosmology can significantly bias the Hubble constant 
and other parameter estimation. However, we demonstrate that a proper 
model independent method such as Gaussian process regression can 
debias the estimation and accurate reconstruct the expansion 
history $H(z)$ including $H_0$.

Furthermore, we illustrate how to use the GP hyperparameters as a 
test to determine whether the data require a beyond-$\Lambda$CDM 
cosmology.  This can be done by fitting a $\Lambda$CDM model to the data, then using this best-fit model as a mean function for the GP regression.  If the posterior of the hyperparameter $\sigma_f$ prefers values significantly different from 0, then that implies the data requires an explanation beyond the best-fit $\Lambda$CDM.  For a ``Next Next Generation'' GW siren dataset, coupled with ``Pantheon-like'' and ``WFIRST-like'' supernovae datasets, GP regression was able to show mock data from reasonable $w_0$--$w_a$ cosmologies was incompatible with the best-fit $\Lambda$CDM cosmology, while accurately recovering the best-fit $\Lambda$CDM cosmology from $\Lambda$CDM-generated mock data.

This could also be used to detect unrecognized systematics in a 
dataset. Using the best fit expansion history from one data set 
(e.g.\ SN) as seed for the GW GP, one can again look for consistency 
with $\sigma_f=0$. 

A particular example of such a systematic could be redshift 
inaccuracy through indirect estimation of the dark siren redshift. We derived constraints on additive and multiplicative systematics, 
showing that even an apparently modest single parameter bias in a 
model dependent fit can actually lead to quite strong bias in joint 
parameter confidence contours. To remove the bias requires the 
additive and multiplicative redshift systematics to be controlled 
at the spectroscopic precision level. 

Cosmic GW sirens alone, even from next next generation 
surveys, will only determine $H_0$ to the $\sim2\%$ accuracy level, 
using the model independent formalism to debias. However, the 
Hubble constant and expansion history can potentially be mapped more 
accurately by using them in conjunction with supernovae and/or 
local GW sirens, with systematics appropriately controlled. These results are not unique to GW sirens and would be applicable to any distance-redshift dataset.

These results necessarily depend on the assumptions built in to the construction of our mock datasets.  Using larger uncertainties on the luminosity distances (especially for those at high redshift) as in \cite{ZhaoWen18}, our results would become less significant.  Either in the reconstruction of the expansion history or in the posterior of the hyperparameters of the GP regression, any deviation away from $\Lambda$CDM would become less significant.

The GP regression code used for this study is made publicly 
available.

\section*{Acknowledgements}

We thank the CosKASI 2019 conference ``The Correlated Universe'' for providing a collaborative venue, and Tamara Davis for discussions about redshift errors and 
$H_0$. A.S. would like to acknowledge the support of the National Research Foundation of Korea (NRF- 2016R1C1B2016478). A.S. would like to acknowledge the support of the Korea Institute for Advanced Study (KIAS) grant funded by the Korea government. 
BL would like to acknowledge the support of the National Research Foundation of Korea (NRF-2019R1I1A1A01063740). 
This work is supported in part by the Energetic Cosmos Laboratory and by 
the U.S.\ Department of Energy, Office of Science, Office of High Energy 
Physics, under Award DE-SC-0007867 and contract no.\ DE-AC02-05CH11231.




\bibliographystyle{mnras}
\bibliography{example} 








\bsp	
\label{lastpage}
\end{document}